\documentclass[12pt,preprint]{aastex}
\slugcomment{Accepted for publication 
in {\it the Astrophysical Journal}  July 22, 2003} 
\def\lax {\ifmmode{_<\atop^{\sim}}\else{${_<\atop^{\sim}}$}\fi} 
\def\gax {\ifmmode{_>\atop^{\sim}}\else{${_>\atop^{\sim}}$}\fi} 

\begin{document}

\title{Broad redshifted line as a signature of   outflow
 }

\author{Lev Titarchuk\altaffilmark{1,2}, Demos Kazanas\altaffilmark{2} and
Peter A. Becker \altaffilmark{3}}

\altaffiltext{1}{George Mason University/Center for Earth
Observing and Space Research, Fairfax, VA 22030; and US Naval Research
Laboratory, Code 7655, Washington, DC 20375-5352; lev@xip.nrl.navy.mil }
\altaffiltext{2}{NASA/ Goddard Space Flight Center, code 660, Greenbelt 
MD 20771, Demos.Kazanas-1@nasa.gov}
\altaffiltext{3}{George Mason University/Center for Earth
Observing and Space Research, Fairfax, VA 22030; pbecker@osf1.gmu}

\begin{abstract}
We formulate and solve the diffusion problem of line photon propagation
in a  bulk  outflow from a compact object (black hole or neutron 
star) using a generic assumption regarding the distribution of line 
photons within the outflow. Thomson scattering of the line photons
within the expanding flow leads to a decrease of their energy which is of 
first order in $v/c$, where $v$ is the outflow velocity and $c$ is
the speed of light. We demonstrate that the emergent line profile is closely 
related to the time distribution of photons diffusing through the flow
(the light curve) and consists of a broad redshifted feature. 
We analyzed the line profiles for the general case 
of outflow density distribution.  We emphasize that the redshifted lines are
intrinsic properties of the powerful outflow that are supposed to be in many
compact objects.

\end{abstract}

\section{Introduction}
The problem of photon propagation in an optically thick fluid in bulk motion 
has been studied in detail in a number of papers. The idea that photons may 
change their energy in repeated scatterings with cold electrons in a moving 
fluid was suggested more than 20 years ago by Payne \& Blandford (1981), 
hereafter PB81 and Cowsik \& Lee (1982). This process, often referred to 
as dynamical or bulk flow Comptonization, is similar in many ways to 
Comptonization by hot electrons once the thermal velocity is replaced by bulk
velocity $\bf v$; there is however a qualitative difference, in that their 
energy gain is linear in the velocity $v$, rather than quadratic as is in
the case of Compton scattering in a medium at rest. Photons diffusing 
and scattering in a medium with bulk flow can gain or lose energy to the 
flow depending on the divergence of its velocity field:  they gain energy 
for $\nabla\cdot {\bf v}<0$, while they lose energy in the opposite case.
The case of a converging radial inflow, where $\nabla\cdot {\bf v}<0$, 
has been treated by (among others) PB81, Nobili, Turolla, \& Zampieri 1993, 
Turolla et al. (1996) and 
Titarchuk, Mastichiadis \& Kylafis (1996; 1997), (hereafter TMK96, TMK97
respectively). These investigations showed that, when monochromatic 
radiation with $\nu = \nu_*$ is injected at large Thomson depth in a 
spherical inflow, the emergent spectrum develops a broad, power-law tail 
at $\nu > \nu_*$. The power law index is related to a combination of the 
flow Thomson depth and its velocity gradient, and this has been invoked to 
explain 
the high energy spectra of accreting BH candidate sources in their ``high" state 
(Laurent \& Titarchuk 1999; Shrader \& Titarchuk 1999).

In this  Paper we treat the problem of an outflow and show that a similar 
broad spectrum is formed but now at energies $\nu<\nu_*$, with the power law 
index dependent on  the velocity gradient. This problem is also relevant 
observationally because of the well known presence of red wings in the 
fluorescent Fe K$\alpha$ line profiles observed in the AGN and galactic 
BH  spectra. 
Turolla et al. (1996) were first who analyzed the radiative transfer in an
expanding atmosphere.
They found a particular solution of the 
PB81 equation for diverging flow using Fourier transformation.
They showed that the overall behavior of the spectra is similar to that of the
converging flow but somehow reversed, since now photons can drift only to
frequencies lower than $\nu_*$.
Furthermore, they  found that adiabatic expansion produces a
drift of injected monochromatic photons towards lower frequencies and the
formation of a power-law, low energy tail with spectral index 3.

Photons can also lose energy in scattering with cold electrons due to recoil. 
This process was treated by Sunyaev \& Titarchuk (1980; hereafter ST80),  
who demonstrated that the emergent spectrum of the high energy photons 
injected  in the static cold electron cloud is determined by the initial 
photon distribution in radius, which in turn dictates the photon escape 
probability distribution in time. The resulting spectrum for monochromatic 
injection is a redshifted feature of spectral width $\Delta\nu $ which 
depends on the line frequency $\nu_*$  and the Thomson depth $\tau_0$, i.e.
$\Delta\nu \sim-\nu_*(h\nu_*/m_ec^2)\tau_0^2$.  
The observation of red-skewed K${\alpha}-$lines observed in a number of AGN 
(and galactic black hole systems) (Fabian et al. 1995), led to the consideration
of the role of this process in producing the observed Fe line profiles.
This was however dismissed by Fabian et al. (1995), 
who correctly pointed out that the
required Thomson downscattering of the line photons would also introduce 
a (non-observed) break in the spectrum at energy $E \simeq m_ec^2/\tau_0^2 
\simeq 20$ keV. An approach along the same lines by Misra \& Kembhavi (1998) 
was rebutted on the basis of line and continuum variability observations
by  Reynolds \& Wilms (2000). 

Following these works, the prevailing process for producing the observed 
profile of the Fe K$\alpha-$line was considered to be Doppler broadening by
the kinematics of a cold ($kT<0.25$ keV) accretion disk reprocessing
the X-ray radiation of an overlying corona into fluorescent Fe K$\alpha-$line
and reflection continuum.  
It was shown (Fabian et al. 1989) that such an arrangment could produce lines of
the desired width and asymmetry provided that the disk extended to its
innermost stable orbit (3 Schwarzschild radii).
 However plausible, this interpretation is 
not entirely without problems; for instance, the (occasionaly) exceedingly 
broad red wing of the line observed requires (in such cases) that the disk 
illumination be concentrated very close to its inner edge 
($F_x\propto r^{-8}$; Nandra et al. 1999) more than most models would allow;
this led Reynolds \& Begelman (1997) to consider fluorescent emission from
within the innermost stable orbit, by matter inspiraling into the black hole. 
Also, as indicated by the models of Nayakshin et al. (2000); Ballantyne et al. 
(2001), the ionization of such a disk by the intense X-ray radiation  might 
invalidate some of basic assumptions associated with this  interpretation. 
 Of additional interest is also the fact that there is a 
marked absence of a blue-shifted wing in the line, a feature expected for a
random orientation of accretion disks and the observers' lines of sight. 
 
Motivated by the above facts and the importance of the Fe K$\alpha-$line in
probing the strong field limit gravity we believe it is important that all
potential alternatives to line broadening be considered in detail. To this
end we examine in the present Paper under what conditions these broad Fe 
line profiles could be also attributed to the effects of an outflow rather 
than solely to accretion disk kinematics. The formulation of the photon 
diffusion problem in an outflow and its solutions are described  in \S 2,
and in \S 3 our results are discussed and certain conclusions are drawn. 

\section{Radiative Transfer in a Bulk Outflow}

\subsection{Main equations and  parameters}

Let $N(r)=N_0(r_0/r)^{\beta}$ be the radial number density profile of an
outflow and let its radial outward speed be
\begin{equation}
v_b/c=(\dot M_{out}/4\pi cN_0r_0^2)(r_0/r)^{2-\beta}=
\dot m_{out} (r_0/r)^{2-\beta}
\end{equation}
obtained from mass conservation in a spherical geometry (here $\dot M_{out}
=4\pi r^2v_bN$). The Thomson optical depth of the flow from some radius r to
infinity is given by
\begin{equation}
\tau=\int_r^{\infty}N_e(r)\sigma_{\rm T}~dr=\sigma_{\rm T}N_0r_0(r_0
/r)^{\beta-1}/(\beta-1),
\end{equation}
where $N_e(r)=N(r)$ is the electron density, $\sigma_{\rm T}$ is the Thomson cross 
section, $r_0$ is a radius at the base of the outflow (this definition 
of optical depth makes the tacit assumption that $\beta>1$) and $\tau_{T,0}=\tau(r_0)=
\sigma_{\rm T}N_0r_0/(\beta-1)$.


The transfer of radiation within the flow in space and energy is governed by 
the photon kinetic equation (BP81, Eq. 18) for the photon occupation
number $n(r,\nu)$, which in steady state reads
%
%
\begin{equation}
-{\bf v_b}\cdot\nabla n+\frac{1}{3}\nabla\cdot(\frac{c}{\kappa}\nabla n)
+\frac{1}{3}(\nabla\cdot{\bf v_b})\,
\nu \frac{\partial n}{\partial \nu}=-{\tilde j}(r,\nu),
\label{radtrans}
\end{equation}
where $\kappa=N_e(r)\sigma_{\rm T}$ is the inverse of the scattering mean free 
path, ${\bf v_b}=v_b{\bf e_r}$, is the flow velocity,  ${\bf e_r}$ is the radial 
unit vector and $\tilde j(r,\nu)$ is the photon source term. 

The spectral flux $F(r,\nu)$ (PB81) is given in terms of $n(r,\nu)$ by 
\begin{equation}
F(r,\nu)= -\frac{1}{3\kappa(r)}\nabla n-\frac{1}{3}v_b \, \nu
\frac{\partial n}{\partial \nu},
\end{equation}
and must satisfy the following boundary conditions: (a) Conservation of 
the frequency integrated flux over the outer boundary, namely 
\begin{equation}
\int F(r,\nu)d\nu\propto r^{-2} ~~~{\rm as}~~ {\rm r\rightarrow \infty}.
\end{equation}
(b) The occupation number should be zero at the inner boundary $r=r_0$, 
where the bulk outflow  starts and the outflow radiation is zero, i.e. 
\begin{equation}
n(r_0)=0.
\end{equation}
%
\subsection{Solution for a general case of any optical depth $\tau_0$} 
According to a theorem (Titarchuk 1994, appendix A), the solution of any 
equation whose LHS operator acting on the unknown function  $n(r,\nu)$
is the sum of two operators $L_{r}$ and $L_{\nu}$, which depend
correspondingly only on space and energy and the RHS, $j(r,\nu)$,
is factorizable, i.e. 
\begin{equation}
L_rn+L_{\nu}n=-j(r,\nu)=-f(r)\varphi(\nu)
\end{equation} 
with boundary conditions independent of the energy $\nu$, i.e.
\begin{equation}
L^{(1)}_rn=0~~~~{\rm as}~~{\rm r\rightarrow \infty}, ~~~~ 
L^{(2)}_rn=0~~~~{\rm for}~~{\rm r=r_0}~, 
\end{equation}
is given by the convolution of the solutions of the time-dependent problem 
of each operator, namely
\begin{equation}
n(r,\nu)=\int_0^{\infty}P(r,u)X(\nu,u)du. 
\label{convol}
\end{equation}
Above $u=N_0\sigma_{\rm T}ct$  is the dimensionless time and 
$P(r,u)$  is the solution of the initial value problem of the 
spatial operator $L_{r}$
\begin{equation}
\frac{\partial P}{\partial u}=L_rP, ~~~~~P(r,0)=f(r)
\label{spaceq}
\end{equation}
with boundary conditions
\begin{equation}
L^{(1)}_rP=0~~~~{\rm as}~~{\rm r\rightarrow \infty},~~~~
L^{(2)}_rP=0~~~~{\rm at}~~{\rm r=r_0},
\label{spacebc}
\end{equation}
%
and $X(\nu,u)$ the solution of the initial value problem of the energy
operator  $L_{\nu}$
\begin{equation}
\frac{\partial X}{\partial u}=L_{\nu}X, ~~~~~X(\nu,0)=\varphi(\nu)
\label{energyeq}
\end{equation}
%
with  boundary conditions
\begin{equation}
\nu^3X\rightarrow 0~~~~{\rm when}~~~\nu\rightarrow 0,~\infty.
\label{energybc}
\end{equation}

In the case where the photon energy change is due to 
the electron recoil (see details in ST80), we have 
\begin{equation}
L_rn=\frac{1}{3\kappa}\nabla\cdot\left[\frac{1}{\kappa}\nabla n\right]
\end{equation}
and
\begin{equation}
L_{\nu}n=\frac{1}{z^2}\frac{\partial}{\partial z}(z^4n).
\end{equation}
where $z=h\nu/m_ec^2$.
The solution of the time-dependent problem for the energy operator $L_\nu$ (the Green's function)
and  $\varphi(z)=\delta(z-z_{\ast})/z^3$  is (see ST80)
\begin{equation}
J(z,u)=z^3X(z,u)=\left\{\begin{array}{ll}
\delta[(1/z-u)^{-1}-z_{\ast}]/z & \mbox{if $z\leq z_{\ast}~{\rm and}~
0\leq u\leq1/z$}\\
0 & \mbox {otherwise,}
\end{array}
\right.\
\label{Jzu}
\end{equation}
where $J(z,u)=z^3X(z,u)$ is the radiation intensity. 
Substitution of Eq.(\ref{Jzu}) into Eq.(\ref{convol}) gives the form of 
the response function (the Green's function) to monochromatic line 
injection in a static cloud
\begin{equation}
G_{z_{\ast}}(z)=z^3n(r_{out},z)=\frac{1}{z_{\ast}z}P[r_{out}, 
(1/z-1/z_{\ast})].
\end{equation}
For a local photon injection the time-dependent diffusion distribution in a
bounded medium has the form of a fast rise and exponential decay 
(see ST80 and Wood et al. 2001), where the characteristic dimensionless time
($u=\sigma_{\rm T}N_0ct$) for the exponential decay is always proportional to 
$\tau_{\rm T,0}^2$, independent of the source distribution. Thus the asymptotic 
form of the redshifted wing of the downscattering line (i.e. for $\nu\ll\nu_*$)
is always exponential, $G_{z_{\ast}}(z)\propto z^{-1}\exp(-A_0/z\tau_{\rm T,0}^2)$.
It is evident from this formula that a significant line  redshift can occur 
only when $h\nu/m_ec^2\tau_{\rm T,0}^2\sim A_0$ where $A_0$ is of order of a few.

To obtain the Green's function of the same problem for a diverging flow 
we use the same method applied in the downscattering case. The time-dependent 
problem for the energy Green's function is (see Eq.\ref{radtrans}
and Eqs. \ref{energyeq}-\ref{energybc})
\begin{equation}
\frac{\partial X}{\partial u}=L_{\nu}X=\nu \frac{\partial X}{\partial \nu}
\label{energyeq2}
\end{equation}
\begin{equation}
X(\nu,0)=\delta(\nu-\nu_{\ast})/\nu^3
\label{energyeqin}
\end{equation}
with  boundary conditions
\begin{equation}
\nu^3X\rightarrow 0~~~~{\rm when}~~~\nu\rightarrow 0,~\infty.
\label{energybc2}
\end{equation}
 The problem (\ref{energyeq2}-\ref{energybc2}) is an initial value problem 
for the first order partial differential equation and  it can be found 
using the method of characteristics (see e.g. Zel'dovich \& Raizer, 2002)
Because the occupation number $X(\nu,0)$ is conserved along the characteristics
$u=\ln\nu_0/\nu$ of equation (18) the solution of the problem 
(18-20) is
\begin{equation}
J_{\nu_{\ast}}(\nu,u)=\nu^3X(\nu,u)=e^{-3u}\delta(\nu e^{u}-\nu_{\ast}).
\label{Jnuu}
\end{equation} 
Substitution of $J_{\nu_{\ast}}(\nu,u)$ from Eq.(\ref{Jnuu}) into Eq.(\ref{convol})
gives us the 
Green's function
\begin{equation}
G_{\nu_{\ast}}(\nu)=\nu^3n(r_{out}, \nu )=\frac{1}{\nu_\ast}\left(\frac{\nu}{\nu_\ast}\right)^3
P[r_{out},\ln(\nu_\ast/\nu)]. 
\end{equation}
The time distrubition $P(r_{out},u)$ can be found from the solution 
of the time-dependent problem of the space operator and initial condition
(see Eq. 10)
\begin{equation}
\frac{\partial P}{\partial u}=L_rP= \frac{\nabla\cdot[(1/\kappa)\nabla P]}
{\nabla\cdot({\bf v_b}/c)}
-3\frac{({\bf v_b}/c)\cdot\nabla P}{\nabla\cdot({\bf v_b}/c)},
\label{diffusion}
\end{equation}
\begin{equation}
P(r,0)=f(r)
\label{spaceinv2}
\end{equation}
with two boundary conditions
\begin{equation}
-(1/\kappa)\nabla P\propto r^{-2}~~~~{\rm at}~~{\rm r\rightarrow \infty},~~~~
P=0~~~~{\rm at}~~{\rm r=r_0}.
\label{spacebc2}
\end{equation}
Introducing $\tau=\tau_{\rm T}$ as a spatial variable Eq.(\ref{diffusion})
can be rewritten as follows
\begin{equation}
\frac{\partial P}{\partial u}= L_rP=\frac{\beta-1}{\beta}
\left\{C_{v_b}^{-1}\tau^{\gamma}
\left[\frac{\partial^2P}{\partial\tau^2}-\frac{2}{(\beta-1)\tau}\frac{\partial P}
{\partial \tau}\right]
+3\tau\frac{\partial P}{\partial \tau}\right\},
\label{diffusion1}
\end{equation}
where $C_{v_b}
=(\dot M_{out}/4\pi cN_0r_0^2)\tau_0^{(\beta-2)/(\beta-1)}=
\dot m_{out}\tau_0^{(\beta-2)/(\beta-1)}$, 
$\gamma=(2\beta-3)/(\beta-1)$.
In order to solve the above time-dependent problem 
we look for solution of the eigenvalue problem for the operator $L_r$ 
(Eq. \ref{diffusion1}) with the appropriate boundary conditions, namely
\begin{equation}
L_r R+\lambda^2 R=0,
\label{eigeneq}
\end{equation} 
with the boundary conditions
\begin{equation}
dR/d\tau\propto \tau^{2/(\beta-1)}\propto r^{-2}~~{\rm as}~~
{\tau\rightarrow0}~~({\rm or}~~ r\rightarrow \infty) ~~~{\rm and}~~~
R=0~~{\rm at}~~{\rm \tau=\tau_0}.
\label{eigenbc}
\end{equation}
The nontrivial solution of equation (\ref{eigeneq}) satisfying the 
boundary conditions (\ref{eigenbc}) (i.e. eigenfuction) is
\begin{equation}
R_{\lambda}(\tau)=\tau^{(\beta+1)/(\beta-1)}\Phi[\beta+1+\lambda^2\beta/3,
\beta+2, -3C_{v_b}(\beta-1)\tau^{1/(\beta-1)}],
\end{equation}
where $\Phi(a,b,z)$ is the confluent (or degenerate) hypergeometric function. 
The boundary conditions (\ref{eigenbc}), along with formula (29), implies 
that the eigenvalues are roots of the equation
\begin{equation}
\Phi(\beta+1+\lambda^2\beta/3,\beta+2,-\tau_{eff,0})=0,
\end{equation}
where $\tau_{eff,0}=3\dot m_{out}(\beta-1)\tau_0$.
Using the asymptotic form of the confluent  hypergeometric function 
 $\Phi(a,b,z)$ for large arguments $z\gg1$ (Abramowitz \& Stegun 1970, 
 Eq.[13.1.4])
 \begin{equation}
 \Phi(a,b,z)=\frac{\Gamma(b)}{\Gamma(b-a)}(-z)^{-a}[1+O(|z|^{-1})]
 \end{equation}
and equation (30) we can find the eigenvalues  for $\tau_{eff,0}=3(\beta-1)\dot m_{out}\tau_0\gg1$ as roots of the following equation
\begin{equation}
1/\Gamma(1-\lambda^2\beta/3)=0,
\end{equation} 
where $\Gamma(z)$ is the gamma-function. 
Because $\Gamma(z)$ goes to $\infty$ for nonpositive integers, $z=-(k-1)$, ($k=1,$~2, 3,...), the eigenvalues are
\begin{equation}
\lambda_k^2=3k/\beta.
\end{equation}
If the monochromatic line sources   are distributed according to the
eigenfunction $f(r)=R_{k}(\tau)$ then $P(\tau,u)$ is 
\begin{equation}
P_k(\tau,u)=R_{k}(\tau)\exp(-\lambda_k^2u).
\end{equation}
Thus when  $f(r)\propto R_{k}(\tau)$ 
the Green's function $G_{\nu_\ast}(\nu)$ [see formula (22)] is a power law
\begin{equation}
G_{\nu_\ast}(\nu)\propto\frac{1}{\nu_\ast}\left(\frac{\nu}{\nu_\ast}
\right)^{3+\lambda_k^2}.
\end{equation}
For $\tau_{eff,0}\gg1$ the spectral index 
\begin{equation}
\alpha_k=3+\lambda_k^2=3(1+k/\beta)
\end{equation}
and the least spectral index  $\alpha=\alpha_1=3(1+1/\beta)$. In Figure 1 we 
present numerical calculations of the least photon index $\Gamma=\alpha-1$
using Eq.(30) for various values of $\beta$ and $\tau_{eff,0}$. As it is seen 
from this figure $\Gamma-$values converge to the asymptotic values 
$\Gamma_{as}= 2+3/\beta$  for $\tau_{eff,0}\gg1$.

In the general case of a given source distribution $f(r)$, $m_{out}$, $\tau_0$, 
and $\beta$ one should calculate the eigenvalues using Eq. (30) and the 
expansion coefficients $b_k$ using the following formulae
\begin{equation}
b_k=\int_0^{\tau_0}\psi(\tau)R_k(\tau)f(\tau)d\tau/H_k(\tau_0),
\end{equation}
\begin{equation}
H_k=\int_0^{\tau_0}\psi(\tau)R_k^2(\tau)d\tau, 
\end{equation}
where $\psi(\tau)=\tau^{(1-2\beta)/(\beta-1)}\exp(\xi)$ is a weight 
function,  $\xi=3C_{v_b}(\beta-1)\tau^{1/(\beta-1)}$ is inversely proportional to the radius $r$
(see Eq. 2) and $C_{v_b} =\dot m_{out}\tau_0^{(\beta-2)/(\beta-1)}$. 
It is worth noting that the variable $\xi$ is equal to the trapping radius for
advection, $r_t$, divided by the radius $r$, where $r_t = (GM/c^2)(\dot M_{out}/\dot M_E)$, 
with $\dot{M}_E = 4 \pi G M m_H /c\sigma_T$ denoting the
Eddington accretion rate (see e.g. Becker \& Begelman 1986,  TMK97).

Equations (37-38) are obtained using the orthogonality properties 
of the eigenfunctions 
$R_k(\tau)$. In appendix A we show that the square norm $H_k$ can be computed analytically in closed form, namely that
\begin{equation}
H_k=p(\tau_0)\frac{\partial R(\lambda_k^2, \tau_0)}{\partial (\lambda^2)}
\frac{\partial R_k(\tau_0)}{\partial \tau},
\end{equation} 
where $p(\tau)=[(\beta-1)/\beta C_{v_b}]\tau^{\gamma}\psi(\tau)$ 
[see definition of $\gamma$ in Eq. (26) above]. Convenient formulae for calculations of derivatives of $R_k$ above are given in
Appendix A. The numerator of equation (37) $c_k$ [a scalar product of $f(\tau)$ and $R_k(\tau)$]
can be also calculated analytically in the case of the $\delta-$function source distribution 
[$f(\tau)=\delta(\tau-\tau_{\ast})$, where $\tau_\ast<\tau_0$]:
$c_k= \psi(\tau_\ast)R_k(\tau_\ast)$.

The formula for the flux ${\cal F }(\nu)=[(r/r_0)^2F(r,\nu)] 
|_{r\rightarrow\infty}$ (see Eq. 4) is
\begin{equation}
{\cal F}(\nu)=[\tau^{2/(1-\beta)}F(\tau,\nu)]|_{\tau\rightarrow0}\propto
\frac{1}{\nu_\ast}\left(\frac{\nu}{\nu_\ast}\right)^3\sum_{k=1}^{\infty}b_k
\left(\frac{\nu}{\nu_\ast}\right)^{\lambda_k^2}.
\end{equation}
In order to calculate $R_{k}$ for  $k\gg1$  in equations  (29, 37-38)  
one should use the asymptotic  form of the confluent functions $\Phi(a,b,z)$ 
for large values of $a$ and $z>1$ (Abramowitz \& Stegun 1970, 
 Eq.[13.5.13]).
\begin{equation}
\Phi(a,b,z)=\Gamma(b)e^{z}(Z/4)^{(1/4-b/2)}\pi^{-1/2}\cos{(Z^{1/2}-b\pi/2+\pi/4),} 
\end{equation}
where $Z=(4a-2b)(-z)$.
In Figure 2 we present the results of calculation of the time distribution 
${\cal P}(u)\propto [\tau^{2/(1-\beta)}\partial P/\partial\tau[\tau, u]|_{\tau\rightarrow0}$
using the expression [see Eq. (34) for the definition of $P_k(\tau,u)$] 
\begin{equation}
P(\tau,u)=\sum_{k=1}^{\infty}b_kP_k(\tau,u) 
\end{equation}
and equations (29-30, 37-38). These particular calculations are made 
for a constant velocity wind ($\beta=2$), $\dot m_{out}=0.8$ and 
$\tau_0=2$, using as initial conditions a monochromatic source
[$\varphi(\nu)=\delta(\nu-\nu_{\ast})$]
and  four different spatial  distributions given by  
$f(\tau)=\exp{(\eta\xi)}$, with 
$\eta=0,~1.5, ~2.5$ (i.e. sources increasingly concentrated towards the base of
the flow for increasing $\eta$)  and  $f(\tau)=R_1(\tau)$, 
i.e. a distribution according to the first eigenfunction. 
 
 All spectra are related to $P(\tau, u)$ through the transformation of
Eq. (22). In Figure 3 we show the evolution of the emergent spectra 
as a function of the initial source distribution $f(\tau)$.
These calculations are made using expression (40). 


\subsection{Solution for the asymptotic case  $\tau_0\gg1$} 

Turolla et al.  (1996) hereafter TZZN, analyzed this case in detail 
and they found analytical
solution of this problem using Fourier transformation of the main equation (3).
Below we show the asymptotic of our solution  for $\tau_0\gg1$. 

Because $\beta\lambda_k/3=k$ for $\tau\gg1$ (see formula 33) we can rewrite
formula (29) for the eigenfunction  as follows 
\begin{equation}
R_{k}(\tau)=\tau^{(\beta+1)/(\beta-1)}\Phi[\beta+1+k,
\beta+2, -\xi].
\end{equation}

The confluent hypergeometric function $\Phi(a,b, x)$ assumes the 
transformation [see Gradshteyn \& Ryzhik 2000 (hereafter GR00), 
equation 9.212.1]:
\begin{equation}
\Phi(b-a,b,-x)=e^{-x}\Phi(a,b,x).
\end{equation}
In our case $a=1-k$ and $b=\beta+2$.
Then we can express $R_k$ through  Laguerre polynomial
\begin{equation}
R_{k}(\tau)=\tau^{(\beta+1)/(\beta-1)}e^{-\xi}L_{k-1}^{\beta+1}
(\xi)
\end{equation}
keeping in mind that 
\begin{equation}
\Phi[1-k,\beta+2, \xi]\propto L_{k-1}^{\beta+1}(\xi)
\end{equation}
(see GR00, equation 8.972.1).
We neglect the constant factor in front of the right hand side of
relation (46) to derive formula (45) because the eigenfunctions 
$R_k$ are always  determined within a constant factor. 

Thus the square norm of $R_k$ is:
$$
H_k=\int_0^{\infty}e^{-\xi}\tau^{3/(\beta-1)}[L_{k-1}^{\beta+1}
(\xi)]^2d\tau=(\beta-1)B^{-(\beta+2)}\int_0^{\infty}e^{-\xi}\xi^{\beta+1}
[L_{k-1}^{\beta+1}
(\xi)]^2d\xi.
$$
We transform
the integral above  using the definition of 
$\xi=3C_{v_b}(\beta-1)\tau^{1/(\beta-1)}=B\tau^{1/(\beta-1)}$ and we obtain  
\begin{equation}
H_k=(\beta-1)B^{-(\beta+2)}
\frac{\Gamma(\beta+1+k)}{(k-1)!},
\end{equation}
where we have used equation (7.414.3) from GR00, to evaluate the integral above for case
$n=m$\footnote{It is worth noting that in the fifth edition of Gradshteyn \&
Ryzhik' table of the integrals there is a typo in  formula (7.414.3).}. It is evident that the constant factor in the right hand side of
equation (47) is independent of k and depends on $\beta$ and $C_{v_b}$ only.  

In order to determine the expansion coefficient $b_n$ for the $\delta-$function
source distribution [$f(\tau)=\delta(\tau-\tau_\ast)$] we need to integrate
the numerator in Eq. (37)  in $\tau$ over a small region around 
$\tau=\tau_{\ast}<\tau_0$. Taking into account Eq. (47) this yields
\begin{equation}
b_k\propto(k-1)!/\Gamma(\beta+1+k)L_{k-1}^{\beta+1}(\xi_{\ast})
\end{equation}
where $\xi_{\ast}=3C_{v_b}(\beta-1)\tau^{1/(\beta-1)}_{\ast}$ and we 
have neglected the constant factor in the right hand side of
equation (48) which is independent of k and depends on $\beta$ and $\tau_*$ only.
Our solution for the flux $\cal F$ is therefore given by (see Eq. 40)
\begin{equation}
{\cal F}(\nu)\propto
\frac{1}{\nu_\ast}\left(\frac{\nu}{\nu_\ast}\right)^3\sum_{k=1}^{\infty}
\frac{(k-1)!}{\Gamma(\beta+1+k)}L_{k-1}^{\beta+1}(\xi_{*})
\left(\frac{\nu}{\nu_\ast}\right)^{3k/\beta}.
\end{equation}
In fact, using formula (8.976.1) of GR00, for this series the final, 
closed-formed solution for the Green's function can be written as
\begin{equation}
{\cal
F}(\nu)\propto\frac{1}{\nu_\ast}\left(\frac{\nu}{\nu_\ast}\right)^{3(1+1/\beta)}
\exp\left[-\frac{(\nu/\nu_{\ast})^{3/\beta}\xi_{*}}{1-
(\nu/\nu_{\ast})^{3/\beta}}\right]
\left[1-\left(\frac{\nu}{\nu_{\ast}}\right)^{3/\beta}\right]^{-(\beta+2)},
\end{equation}
where in formula (8.976.1) of GR00 we have taken the limit $y\to0$ and 
used the asymptotic form of the modified Bessel function $I_{\beta+1}(w)=
(w/2)^{\beta+1}/\Gamma(\beta+2)$ for the argument $w\ll1$.
The shape the spectrum (Eq. 50)  is not identical
to that obtained by TZZN. 
In fact, the low-frequency limit of the spectrum is a power law
\begin{equation}
{\cal F}(\nu)\propto\left(\frac{\nu}{\nu_\ast}\right)^{3(1+1/\beta)}
\end{equation}
with the index $\alpha=\alpha_1=3(1+1/\beta)$. The difference between 
our result and that of TZZN can be explained by the different boundary 
conditions in two cases. It is easy to show that
the TZZN solution obtained using the Fourier transform technique 
is a particular solution of Eq. (3) with a reflection inner boundary  
condition, while our solution is related to an absorption boundary 
condition. A reflection boundary condition, by conserving the number of 
photons, as does the pure scattering involved in the process we examine, 
leads, not surprizingly, to a Wien-like spectrum, i.e. to $\alpha=3$. 
The situation is different
with an absorptive boundary which does not conserve the number of photons,
leading to a harder spectrum as the boundary is more likely to absorb the 
photons that scatter longer, i.e. the lower energy ones. 
It is therefore not by chance that the TZZN index is  3 but our index is 
$3(1+1/\beta)$. A similar situation is obtaind for the spectra of converging 
flows. The spectral indices are different 
depending on either reflection or absorption inner boundary conditions assumed.
TMK96 and TMK97 (see also Turolla, Zane \& Titarchuk 2002) find that 
the spectral indices of the power-law $\nu^{-\alpha}$
are close to $0$ and $2$ in the reflection and absorption cases respectively. 


\section{Discussion and Conclusions}

We have presented above an idealized treatment of the radiative transfer of 
monochromatic photons within a divergent flow, a problem of also observational 
interest, in view of the observations of broad, redshifted Fe lines in the 
extragalactic and galactic black hole  candidate spectra. Not surprisingly, 
the diverging flow leads to a 
redshift of the line the photon energy to produce lines with broad, red 
wings, not unlike those of the Fe K$\alpha$ lines observed. In our view, one of 
the more interesting aspects of our proposal and the results of our calculations
is the natural absence of a corresponding blue wing in this feature without 
the need to invoke a specific geometrical arrangement for the emission. 

Jets and outflows as a means for producing the observed broad red wings
of the Fe K$\alpha$ lines have also been considered by Fabian et al. (1995).
However, these authors had a rather different mechanism in mind: excitation
of the line in an optically thin, cold jet by external X-ray illumination.
The redshifted wing in this case would, they correctly argued, ought to 
be effected by the transverse Doppler effect or a gravitational red shift.
We concur with their assesment that such and set-up for producing
the observed line profiles  would require a rather special geometric
arrangement, as even a large (but smaller than $90^{\circ}$) angle of 
the outflow to the observer's line
of sight should result in blue shifted lines which are generally not seen 
(see however Pounds et al. 2001). Our proposed process is different in
that the line is produced in an (effectively) optically thick medium.
Its red wing is the result of multiple scattering and the associated first
order redshift in each such scattering. This process produces a red wing
to the line without a particularly fine tuned geometric arrangement.

As indicated by our calculations, the most crucial parameter in determining 
the line shape is the effective scattering depth of the flow $\tau_{eff,0}$ 
and the photon source distribution. Clearly, 
a very extended source distribution allows for a large number of photons 
to be produced at regions of very low  scattering depth (case of 
$\eta = 0$ in Figure 3) leading to a rather narrow line profile. An increase 
in the value of $\eta$ (i.e. producing a larger fraction of the photons 
deep into the flow) leads to much broader profiles which can even exhibit 
a shoulder or a secondary minimum, not unlike the Fe K$\alpha$ line feature 
that was interpreted as evidence for infalling matter in NGC 3516 (Nandra et 
al. 1999). The origin of this feature is well understood within our model: 
the line narrow core corresponds to photons produced at low values of 
$\xi$ (large distances $r$), which escape  without much scattering 
in the flow, while the broader, redshifted profile is formed by the photons
(trapped in the flow) 
produced at high values of $\xi$ which diffuse their way out through
the expanding flow. This behavior can be guessed from Fig. 2 where the 
time distribution of photons escaping the flow is presented: There is a 
narrow peak near zero, representing the photons escaping directly 
(essentially with little energy change) and a broader peak comprising the
photons diffusing through the flow while at the same time suffering 
the energy losses that distribute them on the red wing of the line.
It is worthwhile emphasizing that the narrow core would completely 
vanish in the spectrum for the asymptotic case of $\tau_{0}\gg1$ and only 
the broad feature followed by the low frequency power-law are present there. 
We should also point out that photon scattering in diverging flows
introduces besides  redshifted line profiles also negative time 
lags in the spectra (see Figs. 2-3 and formula 22). In fact, a photon 
of initial energy $\nu_*$ loses energy on the way out. If it escapes
after time $u$ ($u$ scatterings) its final energy will be $\nu = 
e^{-u}\nu_*$ and it will escape later (how much that depends on the 
specifics of the scattering plasma) than a photon of energy $\nu_*$
which has not scattered at all. We suggest the combination of these 
features to be intrinsic signatures of any diverging flow.  

Finally, while our results are for simplicity specific to a particular
flow profile, different density and velocity profiles associated
with more general types of outflow, such as ADIOS (Blandford \& Belgelman
1999), could in fact offer much broader parameter space to explore
with more diverse results. We defer such an exploration to a future
publication.

L.T. acknowledges the support of this work by the Center for Earth Observing 
an Space Research of George Mason University. D.K. would like
to acknowledge support by a $Chandra$ GO grant and would like to
thank Frank C. Jones for helpful discussions. We also acknowledge the thorough analysis 
of this paper by the referee and his/her constructive and interesting suggestions.

\appendix

\section{ Orthogonality of the eigenfinctions $R_k(\tau)$ and 
analytical evaluation of the square norm $H_k$ }
We remind the reader the procedure for proving the orthogonality 
of eigenfunctions $R_k$
related to different eigenvalues $\lambda_k^2$ of the boundary problem (27-28). 
One can check that equation (28) can be written in the self-adjoint form
\begin{equation}
\frac{d}{d\tau}\left(p(\tau)\frac{dR_k}{d\tau}\right)+\lambda_k^2\psi(\tau)R_k=0
\end{equation}
where 
\begin{equation}
p(\tau)=[(\beta-1)/C_{v_b}\beta]\tau^{\gamma}\psi(\tau),
\end{equation} 
and
\begin{equation}
\psi(\tau)=\tau^{(1-2\beta)/(\beta-1)}e^\xi. 
\end{equation}
Let us suppose that $\lambda_n^2$ and $\lambda_m^2$ are eigenvalues corresponding
to the eigenfunctions $R_n(\tau)$ and $R_m^2$ respectively.
Then we use equation (A1) to write 
\begin{equation}
R_m\left\{\frac{d}{d\tau}\left(p(\tau)\frac{dR_n}{d\tau}\right)+\lambda_k^2\psi(\tau)R_n\right\}=0,
\end{equation}
\begin{equation}
R_n\left\{\frac{d}{d\tau}\left(p(\tau)\frac{dR_m}{d\tau}\right)+\lambda_k^2\psi(\tau)R_m\right\}=0.
\end{equation}
Subtracting the second from the first yields, upon integration,
\begin{equation}
(\lambda^2_n -\lambda^2_m)\int_0^{\tau_0}\psi(\tau)R_n(\tau)R_m(\tau)d\tau=
p(\tau)\left[R_n(\tau)\frac{dR_m}{d\tau}-R_m(\tau)\frac{dR_n}{d\tau}\right]\Bigg|^{\tau_{0}}_0.
\end{equation}
Because of homogeneous boundary conditions (28) the right-hand side of equation (A6) vanishes, and
consequently we conclude that 
\begin{equation}
(R_n,R_m)=\int_0^{\tau_0}\psi(\tau)R_n(\tau)R_m(\tau)d\tau=0,~~~~ {\rm if}~~\lambda_n^2\ne\lambda_m^2.
\end{equation}
This result establishes the orthogonality of the family of eigenfunctions $R_k(\tau)$.
Ordinarly, one must resort to numerical integration in order to compute the square norm
$H_k=||R_k||^2$ of the eigenfunctions $R_k(\tau)$. Suppose that $\lambda_n^2$ is the eigenvalue
corresponding to the eigenfunction $R_n(\tau)$, and $\lambda$ is an arbitrary value associated with
the general solution $R(\lambda^2,\tau)$ given by equation (27). Then, in analogy with equation (A6),
we can show that 
\begin{equation}
\int_0^{\tau_0}\psi(\tau)R(\lambda,\tau)R_k(\tau)d\tau=
\frac{p(\tau)\left[R\frac{dR_k}{d\tau}-R_k\frac{dR}{d\tau}\right]
\Bigg|^{\tau_{0}}_0}{(\lambda^2 -\lambda^2_k)}.
\end{equation}
The integral that we seek is obtained in limit $\lambda^2\to\lambda_k^2$. However, in this limit,
both the numerator and denominator in equation (A8) vanish, and therefore 
must use L'H\^opital's
rule to obtain
$$
H_n=||R_k||^2=\lim_{\lambda^2\to\lambda_k^2}\int_0^{\tau_0}\psi(\tau)R(\lambda,\tau)R_k(\tau)d\tau
$$
\begin{equation}
=\lim_{\lambda^2\to\lambda_k^2}p(\tau)\left[\frac{\partial R}
{\partial(\lambda^2)}
\frac{dR_k}{d\tau}-R_k\frac{\partial^2R}{\partial\tau\partial(\lambda^2)}
\right]\Bigg|^{\tau_{0}}_0.
\end{equation}
Because $R_k(\tau)$ goes to zero when $\tau\to0$ and $\tau=\tau_0$ and $dR_k/d\tau\to 0$ when
$\tau\to0$ (see Eq. 28) our expression for $H_k$ reduces to
\begin{equation}
H_k=p(\tau_0)\frac{\partial R(\lambda_k^2,\tau_0)}
{\partial(\lambda^2)}
\frac{dR_k(\tau_0)}{d\tau}.
\end{equation}
Thus  there is no need to employ numerical integration to determine the values of $H_k$.

Using the differential property of the confluent function $\Phi(a,b,x)$ 
(Abramowitz \& Stegan
1070, Eq. 13.4.8) and the condition (29)
we find that 
\begin{equation}
\frac{dR_k(\tau_0)}{d\tau}=-3\dot m_{out}\tau_0^{3/(\beta-1)}
\frac{\beta+1+\lambda_k^2\beta/3}{\beta+2}\Phi[\beta+2+\lambda_k^2\beta/3,\beta+3, 
-3\dot m_{out}(\beta-1)\tau_0].
\end{equation}
The derivative 
\begin{equation}
\frac{\partial R(\lambda_k^2,\tau_0)}
{\partial(\lambda^2)}=\tau_0^{{\beta+1}/(\beta-1)}
(\beta/3)\Phi^{\prime}_a[\beta+1+\lambda_k^2\beta/3,\beta+2, -3\dot m_{out}(\beta-1)\tau_0]
\end{equation}
can be calculated using the function $\Psi(z)$ defined as  $\Psi(z)=\Gamma^{\prime}(z)/\Gamma(z)$. 
In fact, 
\begin{equation}
\Phi^{\prime}_a(a,b,z)=\frac{z}{b}+\frac{(a)_2^{\prime}z^2}{(b)_22!}+
\frac{(a)_3^{\prime}z^3}{(b)_33!} +...+\frac{(a)_n^{\prime}z^n}{(b)_nn!}+..
\end{equation}
where $(a)_n=a(a+1)(a+2)...(a+n-1), ~(a)_0=1$  and
\begin{equation}
(a)_n^{\prime}=(a)_n[\Psi(a+n)-\Psi(a)].
\end{equation}
 One can calculate $(a)_n^{\prime}$ using the  expansion  of $\Psi-$function 
 for large value arguments $z\gg1$ (Stegan \& Abramowitz 1970, Eq. 6.3.18) 
\begin{equation}
\Psi(z)\sim \ln~z-1/2z -1/12z^2+1/120z^4-1/252z^6+... 
\end{equation}



\begin{figure}
\includegraphics[width=7in,height=8in,angle=0]{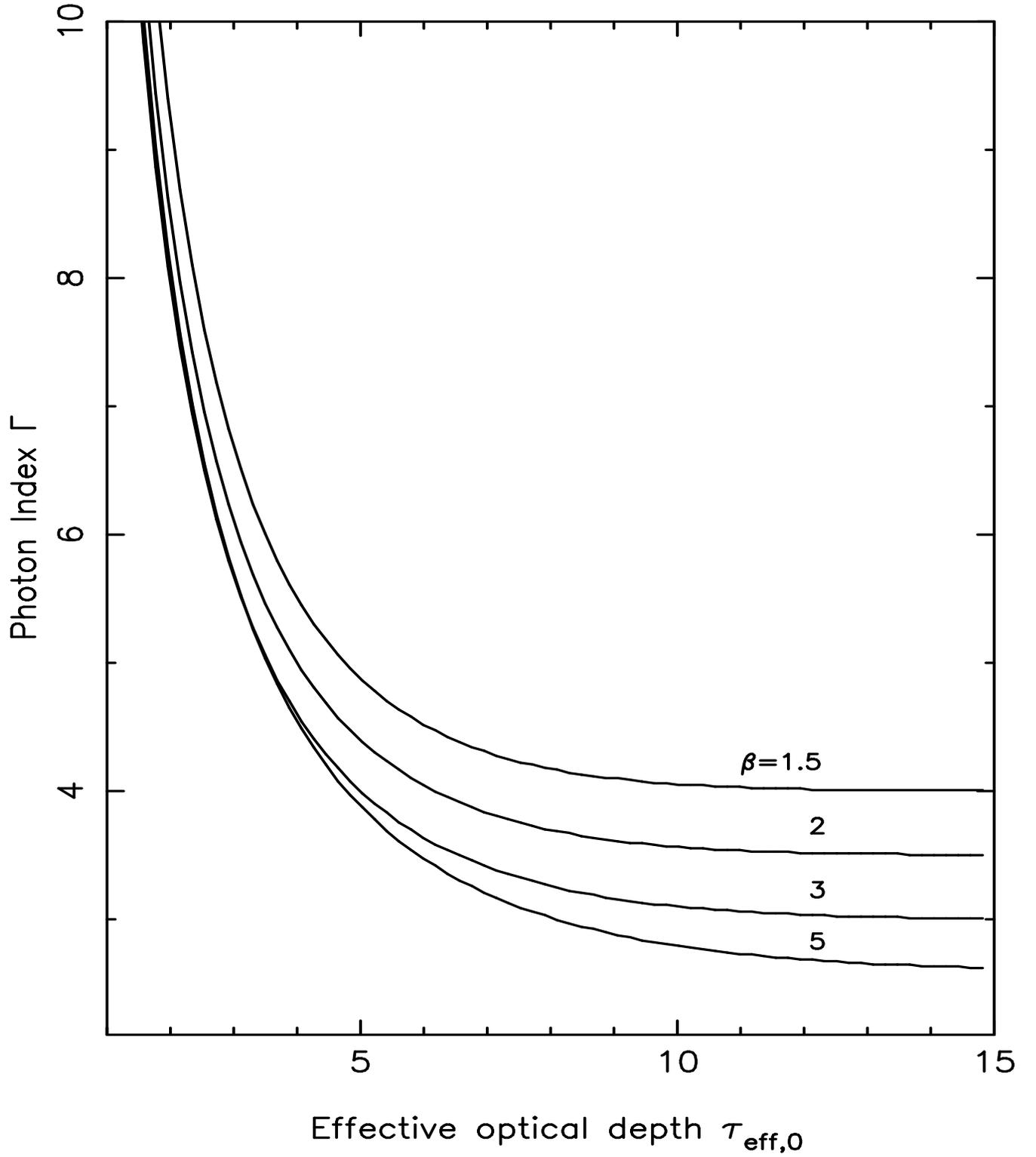}
\caption{Plot of the photon index $\Gamma=\Gamma_1$ vs. effective optical depth 
$\tau_{eff,0}=3(\beta-1)\dot m_{out}\tau_0$ for various values of $\beta$.
 }
\end{figure}
\begin{figure}
\includegraphics[width=7.in,height=6in,angle=-90]{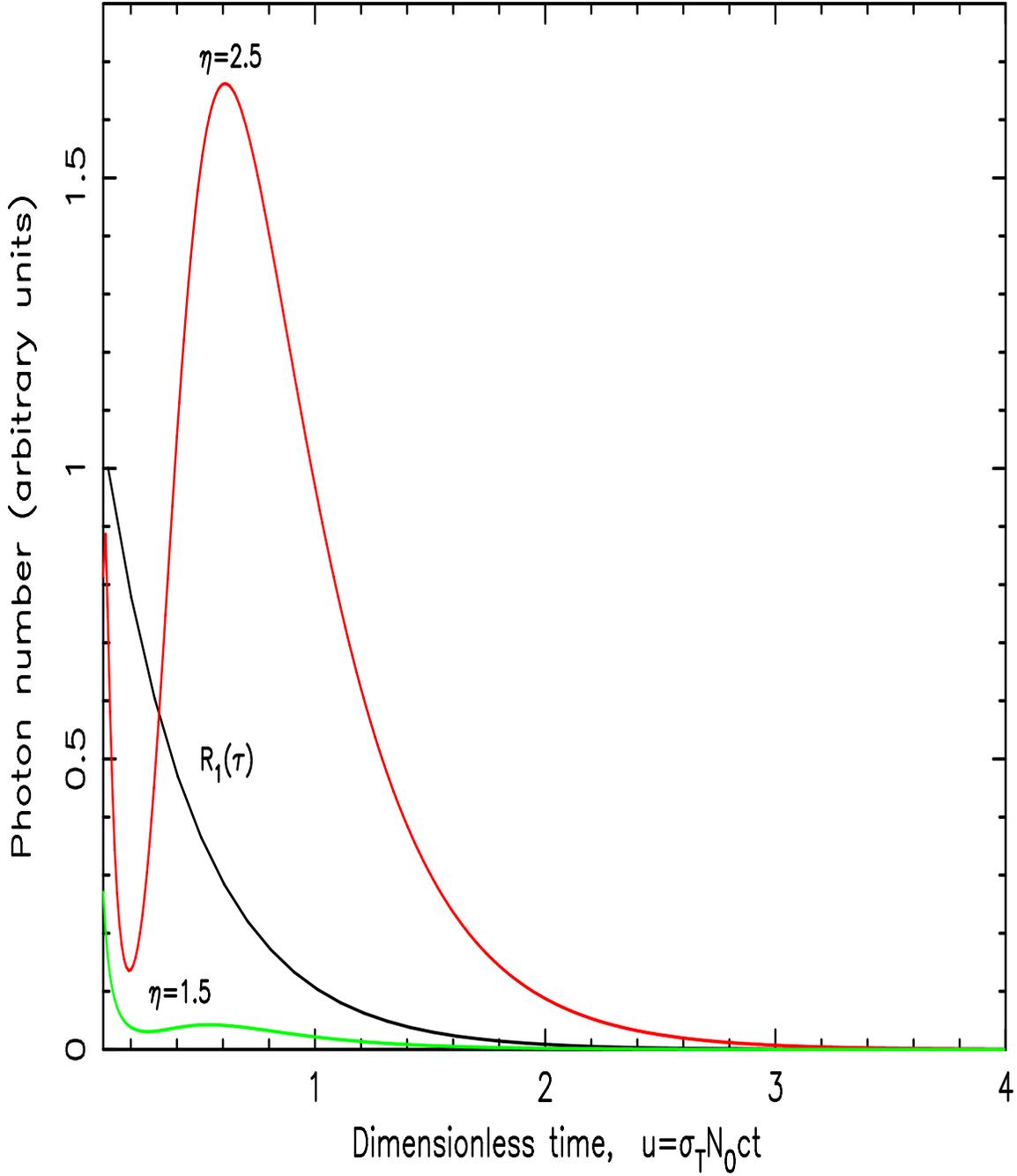}
\caption{Plot of  ${\cal P}(u)$ \{${\cal P}(u)\propto [\tau^{2/(1-\beta)}\partial
P/\partial\tau[\tau, u]|_{\tau\rightarrow0}$\} for three monochromatic source distributions:
$f(\tau)=\exp{(\eta\xi)}$, for $\eta=1.5,~2.5$  and 
distributed according to the first eigenfunction $R_1(\tau)$ (see text). 
Plot for $\eta=0$ is not displayed here, because  it is concentrated very close
to $u=0$ in the presented scale. Calculations are made for $\beta=2$, 
$\dot m_{out}=0.8$ and $\tau_0=2$.
}
\end{figure}
\begin{figure}
\includegraphics[width=7.in,height=6.5in,angle=-90]{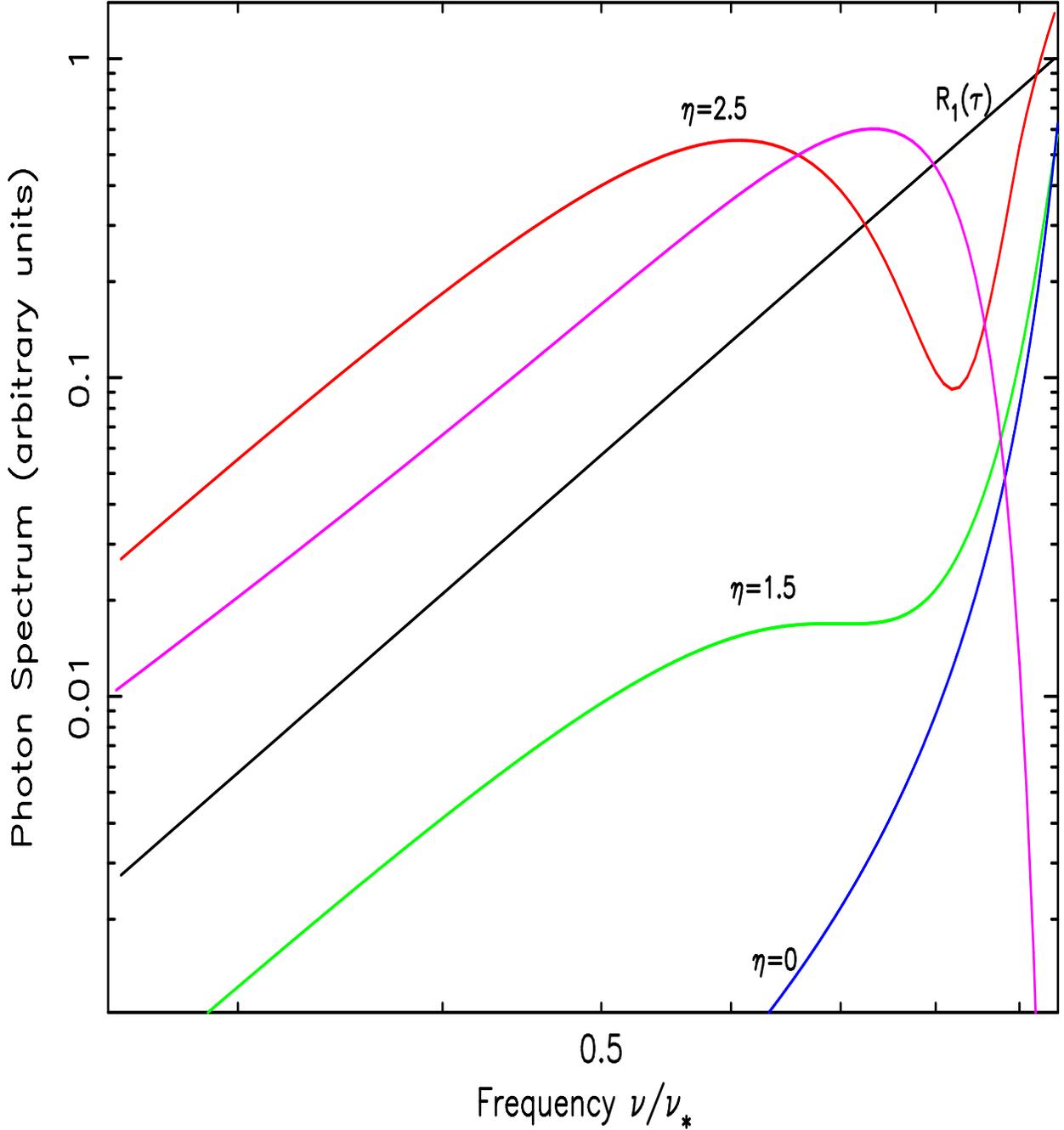}
\caption{Plot of  emergent photon redshifted line spectra. There are  four spectra for
 monochromatic source distributions.
$f(\tau)=\exp{(\eta\xi)}$, for $\eta=0,~1.5,~2.5$  and 
distributed according to the first eigenfunction $R_1(\tau)$ (see text). All these four spectra are
related to ${\cal P}(u)$ through transformation (22) (see text). Calculations are made for $\beta=2$, 
$\dot m_{out}=0.8$ and $\tau_0=2$. The fifth photon spectrum (magenta) is for the asymptotic case 
of $\tau_0\gg1$ (see formula 50). The monochromatic $\delta-$ function source is located at
$\xi=\xi_*=2$, and $\beta=2$.} 
\end{figure}
\


\end{document}